# The effect of magnetic field on the two superconducting gaps in MgB$_2$

Y. Bugoslavsky, Y. Miyoshi, G.K. Perkins, A.D. Caplin, L.F. Cohen

Imperial College London

A. V. Pogrebnyakov, X. X. Xi

Department of Physics and Materials Research Institute, Pennsylvania State University, University Park, PA 16802

Double-gap superconductivity in an epitaxial MgB$_2$ film has been studied by means of point-contact spectroscopy in magnetic field up to 8 Tesla. The relatively fast disappearance of the feature associated with the π-band gap at a field around 1 T is caused by the line broadening due to strong pair breaking rather than to a collapse of the double-gap state. This pair breaking was found to increase linearly with field. Field dependences of the order parameters $\Delta_\pi$ and $\Delta_\sigma$ in π and σ bands was measured in field applied parallel and perpendicular to the film, at T = 4.2 K. In perpendicular field, both order parameters survive to a common $H_{c2}$, which is about 6.5 T for this direction. In parallel field, the decrease of the σ gap is much more gradual, consistent with the $H_{c2}$ being about 4 times greater in this orientation. The difference in $\Delta_\pi$ measured in the two field orientations is however smaller than the difference of $\Delta_\sigma$. We compare these results with the data on tunnelling spectroscopy and specific heat measurements of MgB$_2$ single crystals and find consistency between the different measurements.

There is compelling evidence that magnesium diboride, MgB$_2$, is an unusual superconductor that has two intrinsic order parameters. Soon after the discovery of superconductivity in MgB$_2$, there was rapid progress in understanding the basics of the double-gap structure, both experimentally and theoretically. It is now generally accepted that the smaller gap $\Delta_\pi$ arises from π bands, which have 3-dimensional Fermi surfaces (FS), whereas the larger gap $\Delta_\sigma$ is associated with the σ bands, whose FS are sets of cylinders with axes parallel to $k_z$ and so 2-dimensional [1]. Recent experiments that have probed the electron density of states directionally in single crystals agree with the predicted band symmetries [2,3]. In spite of the remarkable agreement between theory and experiment up to now, some key questions still remain open. It is known that MgB$_2$ is an anisotropic superconductor (the anisotropy of the upper critical field is sample-dependent, but in any case modest, in the range 2-6). It has been suggested [3,4] that, unlike common BCS superconductors, MgB$_2$ has an additional "intermediate" upper critical field $H_{c2\pi}$, at which superconductivity in the π band is suppressed, while the larger σ gap is retained. However, the anisotropy should increase if the 3D component were quenched at $H_{c2\pi}$, but no marked change has been seen, and hence this interpretation is controversial.

We have performed point-contact spectroscopy on an epitaxial MgB$_2$ film, so as to study the evolution of the structure of the order parameter with magnetic field. We find that the small gap does not collapse in small field, but the double-gap state is retained up to the common $H_{c2}$ (which is for the field perpendicular to the film within the reach of our magnet). We ascribe the fast increase of spectral broadening with the field to orbital pair breaking. This interpretation is consistent with other recent experiments on point-contact and scanning tunneling spectroscopy and heat capacity of MgB$_2$ single crystals.

The 100-nm thick MgB$_2$ film was grown using hybrid physical-chemical vapour deposition technique [5] on a sapphire substrate; structural studies of similar films have shown epitaxial growth with the c-axis oriented normal to the film surface [6]. The resistive transition was at 39 K (Fig. 1). Judging by the resistivity at 40 K (ρ ~ 10 μΩ cm) and the upper critical field ($H_{c2}^\parallel$(4.2 K) ~ 6.5 T), the film has stronger intra-band scattering than pure MgB$_2$ crystals, for which the corresponding values are ~ 2 μΩ cm and ~ 3.5 T [7]. Assuming full intergrain connectivity (as it was done in [8]), one can estimate the scattering rates $\gamma_\sigma$ and $\gamma_\pi$ in the two-band model. The calculation suggests that the σ band is in the clean or moderately dirty regime ($\pi\Delta_\sigma \geq \gamma_\sigma$), but the π band is well into the dirty limit ($\pi\Delta_\pi \ll \gamma_\pi$) [9]. The anisotropy of $H_{c2}$ showed a weak temperature dependence (Inset, Fig 1).

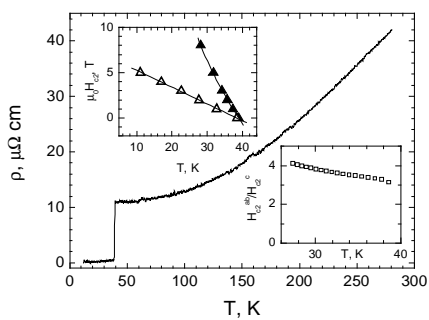

Fig. 1. Resistivity of the MgB$_2$ film versus temperature. Top inset: upper critical field $H_{c2}$ parallel (filled symbols) and perpendicular (open symbols) to the film, as obtained from R(T) curves taken at different fields. Bottom inset: the anisotropy factor, $H_{c2}^\parallel/H_{c2}^\perp$ as a function of temperature

Building on our previous study [10], we have performed point-contact spectroscopy in magnetic fields up to 8 Tesla. The sample can be mounted with the field either parallel ($H^\parallel$) or perpendicular ($H^\perp$) to the film. The field direction is known within a few degrees. The point contact is created using a sharp gold tip, which is perpendicular to the film.

According to the symmetry of the electron bands in $MgB_2$, a directional probe that injects electrons parallel to the crystalline c-axis should predominantly detect a contribution from the 3D π gap. This has been observed in the tunnelling experiments on single crystals [11] and films [2]. The point-contact measurements on single crystals showed however that clean single-gap spectra are rarely seen even for nominal c-axis injection [3]. This fact has been attributed to injection occurring into a broad cone in the point-contact experiment, so that the probe always couples to the 2D σ band as well. In our films likewise, a significant coupling to the σ-gap states is often observed. The relative weight of the σ-gap contribution is higher for junctions with low normal-state resistance $R_0$. Typically, well-resolved double-gap structures are seen when $R_0$ is below 20 Ohm, otherwise only the π gap is resolved (Fig 2 (a) and (b)).

There are additional spectral features at energies outside the gap region (typically 15-30 meV at zero field), so that the reference point for normalising the spectra has to be taken at about 40-50 meV, where the background becomes flat.

The spectra were analysed in the framework of the BTK theory [12], using a sum of appropriately weighted contributions that correspond to the two gaps. There are five adjustable parameters used in the fitting: the gap energies, $\Delta_\pi$ and $\Delta_\sigma$, the dimensionless tunneling barrier Z (assumed to be the same for both gaps), the ratio of the π and σ-gap contributions and the generic smearing parameter ω. The latter accounts for all extrinsic and intrinsic pair-breaking mechanisms. In calculating the model spectra, the smearing effects were taken into account by convolution of the zero-temperature BTK results with a Gaussian of dispersion ω. The Gaussian function is a good approximation to the derivative of the Fermi distribution, which is used to calculate finite-temperature effects [12]. The dispersion is then related to the temperature as $\omega = \sqrt{2}\, T$. There is, however a non-thermal contribution to ω. The smallest smearing observed at 4.2 K (=0.4 meV), was about 1 meV, so that the non-thermal part was of the order of 0.6 meV. Provided ω is small enough for the peaks in the spectra to be clearly resolved, fitting to the BTK model is robust in spite of the large number of variables used.

Variation of the gap values between different junctions was observed. In Fig 2(c) we present the statistics of the measured values compared to the calculated distribution of the gaps over the FS [13]. The calculations were made in the limit of negligible scattering. The strong intra-band scattering must average out the directional dependence of the gaps. As a result of k-space averaging, the gap dispersion should be seen in the actual experiment as a non-thermal contribution to the spectral broadening. Indeed, the value of this broadening is in good agreement with the characteristic width of the theoretical distribution. In the strong intra-band scattering regime, the variation of the measured gaps is likely to be caused by the surface-induced inter-band scattering, which would be sensitive to the details of the junction interface structure, and therefore practically out of control in our experiment. We note however that the experimental distribution is within the bounds of the clean-limit theoretical prediction, indicating that the inter-band scattering is not strong.

Fig 3(a) shows an example of the evolution of the two-gap spectrum with increasing magnetic field applied parallel to the film. The solid lines are the best fits to the BTK model. The apparent effect of the magnetic field is that the central pair of peaks corresponding to the π-band is suppressed whereas the outer σ-band peaks barely change. This observation has to be interpreted with caution though, as the suppression of peaks does not necessarily mean closing of the gap. Indeed, the spectrum at 1 Tesla retains well-defined shoulders, indicative of the smaller gap; the position of the shoulders is almost the same as the position of the zero-field peaks. Irrespectively of the fitting, the peaks positions are good indication of nearly constant values of $\Delta_\pi$ and $\Delta_\sigma$, which decrease only slightly in fields up to 1.5 Tesla. There is no evidence of the collapse of the π gap (Fig 3(b)). The lower two curves in Fig 3(a) correspond to a different

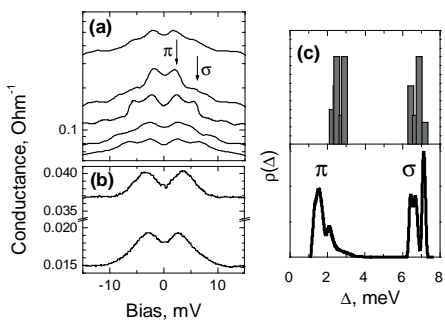

Fig 2. Examples of double-gap (a) and single-gap (b) spectra at 4.2 K, zero field, taken at different locations on the sample. The dashed lines indicate the gap values. (c): comparison of the experimental statistics of the gap values (top) with the theoretical distribution [13]

junction, where there was a negligible σ-gap contribution at zero field (as shown in Fig.2(b)). The presence of a spectral feature is an unambiguous indication of the retention of the π gap up to 5 T, despite limited reliability of the BTK fit at this field. At fields up to 1.5 T all the fit parameters remain nearly constant, except for the line broadening ω, which increases linearly with field at a rate of about 0.5 meV/T (Fig 3(c)). The increasing broadening explains the disappearance of the narrow central peaks from the spectra: the signature of the π gap gets washed out because ω approaches $\Delta_\pi$, whereas the σ gap is less sensitive to broadening simply because of its larger value, so ω is smaller than $\Delta_\sigma$. This interpretation is consistent with the point contact data on $MgB_2$ single crystal [3].

Unlike thermal smearing, the use of Gaussian convolution in fitting the data taken in magnetic field is not *a priori* justified. However, given the fact that the model represents the evolution of the data remarkably well, the inferred parameters should be at least meaningful. The increasing line broadening is indicative of increasing density of single particle excitations induced by the magnetic field. It is worth noting that if spatial averaging of the order parameter over the vortex cores had primary significance (as suggested in [11] and [14]), the measured $\Delta_\pi$ would have been a rapidly decreasing function of H, contrary to our observation. Therefore an alternative interpretation is needed to explain the evolution of point-contact spectra at low fields.

Magnetic pair breaking has been studied in great detail in conventional superconductors, as caused by either spin-flip scattering (which here, in absence of magnetic impurities, is irrelevant) or orbital pair breaking (OPB) [15]. In the dirty limit, OPB is characterised by an energy parameter Γ, which increases proportionally to $H^2$. In the clean limit [16], the calculated zero-temperature, in-field spectra cannot be reproduced by simple smearing out of the zero-field spectrum, therefore a single "smearing parameter" cannot be extracted. However, the qualitative conclusion important for the current discussion is that the degree of pair-breaking at a given field is critically dependent on the mean-free path, with clean superconductors being more strongly affected [16].

It is likely that the presence of considerable thermal broadening masks the details (in particular, asymmetry with respect to the gap edge) of OPB in the spectra. The magnetic field effectively produces an additive contribution to the broadening. It is interesting to note that spectra of Al-Pb junction at 1.74K [17] are well Gaussian-broadened. We have analysed the tunneling data on $MgB_2$ crystal [11], and found that the magnetic broadening is also Gaussian, and increasing linearly with field (Fig 3(c)).

Whether this linear dependence is specific to the double-gap superconductivity has yet to be understood. However, the scattering certainly plays a role in magnetic pairbreaking. As Fig 3c shows, in the single crystal ω increases with field much faster than in our thin film, in agreement with the crystal being closer to the clean limit.

With the field applied perpendicular to the film, the behaviour of the spectra was distinctly different (Fig. 4). The qualitative observation (importantly not relying on the validity of the fit) is that the width of the trapezoidal shape associated with the larger σ gap decreases noticeably faster in perpendicular than in parallel field. The effective barrier Z was different in the two cases. The lower Z in the case $H^\perp$ results in a more shallow dip at zero bias, so that the feature of the π gap appears as a clear peak even in the much-smeared 3 Tesla spectrum. Therefore, in high perpendicular field likewise, the small gap is clearly retained.

The variation of the gaps across the entire field range is summarised in Fig. 5. At fields above 3 T spectral distortions become pronounced and the fit less robust, hence larger error bars. The π gap at high $H^{\parallel}$ was estimated from a single-gap spectrum; the results are shown as open circles in Fig 5. For $H^{\perp}$, both gaps tend to close simultaneously at the bulk $H_{c2}^{\perp}$ of about 6.5 Tesla. In $H^{\parallel}$, the σ gap decreases slower than in $H^{\perp}$, which agrees qualitatively with the estimated $H_{c2}^{\parallel}$ of ~30 T. However, the difference in the π gap behaviour at the two field directions is just beyond the estimated error bars. We suggest that this effect is due to the influence of magnetic vortices on the spectra.

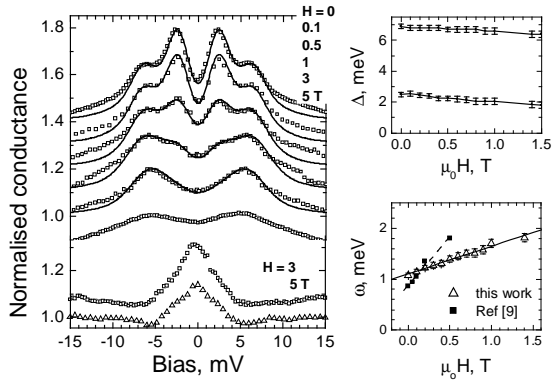

Fig 3. (a) The evolution of the point-contact spectra with increasing field at T=4.2 K, $H^\parallel$. Top to bottom: H=0; 0.1; 0.5; 1, 3 and 5 T. The lower two curves are for a single-gap junction, at H=3 and 5 T. The curves are shifted for clarity. The solid lines are fit to the BTK model. (b) The values of the two gaps as inferred from the fit. (c) Open symbols - the broadening parameter for spectra of panel (a). The solid symbols represents an analogous parameter for the tunneling data of Ref 11

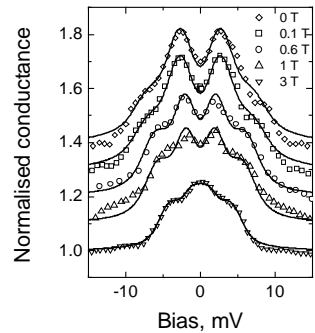

Fig 4. The point contact spectra in perpendicular field.

The fact that the film is in the mixed state has a two-fold consequence. Firstly, the high density of vortices at $H \gg H_{c1}$ ensures that magnetic screening is negligible and the entire volume of the film is penetrated. The OPB is in this case a bulk rather than a surface effect. On the other hand, the size of the point contact (up to 30 μm) is much greater than the vortex separation (100 nm at 0.1 T). Due to the presence of large number of normal cores (wherein Δ is suppressed), the measured average value of Δ will be smaller than the "bulk" Δ (i.e., taken away from the cores), and the amplitude of the conductance feature would be suppressed due to the absence of Andreev reflection in the core regions. The latter effect causes distortion of the spectra and would yield poor fits. In contrast, the fact that we obtain good fits at fields below 1.5 T suggests that the current is primarily injected into the "bulk" areas, and these regions contribute most to the measured spectra. The fitting starts to fail at fields where the vortex cores occupy a substantial fraction of the sample volume, so that the effect of averaging becomes important. In this high-field regime the estimated values of Δ plotted in Fig 5 are then the lower bound of the "bulk" order parameter. This argument can explain the apparent similarity in the of $\Delta_{\pi}$ at two field orientations. It is likely that the decrease of the "bulk" $\Delta_{\pi}$ at $H^{\parallel}$ is more gradual than measured, and therefore consistent with high value of $H_{c2}^{\parallel}$.

Comparison of our point-contact results with the tunneling data [11] suggests that similar mechanisms determine the OPB in MgB$_2$ crystal and epitaxial film. As OPB effectively results in enhanced density of single-particle excitations, it is very likely that it is the dominant cause of the unusual field dependence of the heat capacity in MgB$_2$ single crystals [4]. Further study of the correlation between the OPB and the intra-band scattering rate may prove useful in determining whether the linear increase of the pair breaking parameter is related to the two-gap structure, or if it is a general property of clean systems. It would be also important from the practical point of view, as the strong suppression of the density of the superconducting condensate in magnetic field has a detrimental effect on the critical current density.

In conclusion, the point-contact spectroscopy of MgB$_2$ film in magnetic field has shown that both σ and π-band order parameters are retained at high fields, but the strong orbital magnetic pair breaking causes significant broadening of the spectra. The resulting enhancement of the density of single-particle excitations explains the unusual features observed in the tunneling spectroscopy and heat capacity experiments.

We are grateful to O.V.Dolgov for estimating the scattering rates, Alex Gurevich, A.A.Golubov and I.I.Mazin for useful discussions. This work was supported by the UK Engineering and Physical Sciences Research Council. Work at Penn State was supported in part by ONR under grant No. N00014-00-1-0294.

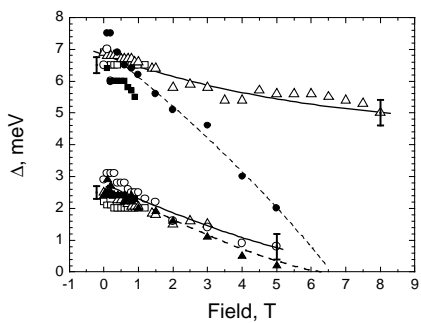

Fig 5. The values of the order parameter obtained from the point-contact spectra. Solid symbols: $H^{\perp}$; open symbols: $H^{\parallel}$. Lines are guide to the eye. The error bars indicate typical uncertainty of the values at low and high fields.